\documentclass[10pt,twocolumn,english]{article}

\usepackage[T1]{fontenc}
\usepackage[latin9]{inputenc}
\usepackage[a4paper]{geometry}
\usepackage{color}
\definecolor{document_fontcolor}{rgb}{0, 0, 0}
\color{document_fontcolor}
\usepackage{amstext}
\usepackage{stackrel}
\usepackage{graphicx}
\usepackage{setspace}
\usepackage{amsmath}
\usepackage{mathtools}
\usepackage{amssymb}
\makeatletter

\newenvironment{lyxlist}[1]
{\begin{list}{}
{\settowidth{\labelwidth}{#1}
 \setlength{\leftmargin}{\labelwidth}
 \addtolength{\leftmargin}{\labelsep}
 }}
{\end{list}}

\@ifundefined{date}{}{\date{}}

\usepackage{array}\usepackage{amsthm}\usepackage{float}

\newcommand{\lyxmathsym}[1]{\ifmmode\begingroup\def\b@ld{bold}
  \text{\ifx\math@version\b@ld\bfseries\fi#1}\endgroup\else#1\fi}

 \@ifundefined{definecolor}{

}{\makeatletter} \AtBeginDocument{
  
}

\usepackage{fancyhdr}\usepackage{titlesec}\titleformat{\chapter}[display]
  {\normalfont\huge\bfseries\centering}{\MakeUppercase\chaptertitlename\ \thechapter}{20pt}{\LARGE}
\usepackage[square, numbers, comma, sort&compress]{natbib}

\usepackage{babel}

\makeatother

\usepackage{babel}
\begin{document}

\title{Prepare-and-measure based QKD protocol under free-space losses }

\author{Mitali Sisodia\thanks{\emph{Email: mitalisisodiyadc@gmail.com}}
and Joyee Ghosh\thanks{\emph{Email: joyee@physics.iitd.ac.in}}\\
Department of Physics, Indian Institute of Technology Delhi, \\
New Delhi 110016, India }
\maketitle
\begin{abstract}
In this study, we have theoretically presented prepare-and-measure
based SARG04 protocol over free-space. It has shown that the highest
secret key rate is possible even under free-space losses with a maximum
tolerance of noise. 
\end{abstract}
\begin{lyxlist}{00.00.0000}
\item [{\textbf{Keywords:}}] Quantum key distribution, Secure key rate,
Geometrical losses, Atmospheric losses. 
\end{lyxlist}

\section{Introduction}

Quantum key distribution (QKD) allows to two parties (Alice (sender)
and Bob (receiver)) to generate unconditional secure keys in the absence
or presence of Eve and noises. Bennett et al., proposed the first
theoretical QKD model and the first experimental implementation on
it for 32cm channel length in 1984 \cite{BB84} and 1989 \cite{EXbb},
respectively. Since then, tremendous progress of QKD protocols have
been made such as, B92 \cite{B92}, BBM92 \cite{BBM}, Differential
phase sift \cite{dps}, coherent-one way \cite{COW} etc. It has also
been proved that the implementation of QKD protocol is possible globally
(satellite-based) such as, satellite-based decoy-state QKD is implemented
for 1120 km \cite{1120}, a group of Ursin implemented QKD protocol
for 144km \cite{Ursin}, and so on.

BB84 is the most common widely used QKD protocol, which is unconditionally
secure with the single-photon states. However, in the practical conditions,
highly efficient single-photon sources is not available, an alternate
of single-photon source, i.e. strongly attenuated laser pulse source
has been used by many researchers to perform secure quantum cryptography.
The pulsed laser source generates weak coherent state, and the photon
number of each pulse follows Poisson distribution $\stackrel[n=0]{\infty}{\sum}\frac{\mu^{n}}{n!}e^{-\mu}|n\rangle\langle n|$,
whereas $n$ represents the $n$-photon state, $\mu$ is the average
photon number ($\mu\prec1$) \cite{Source} . Weak coherent pulses
can be contained one photon, two-photon or multi-photon in a pulse
and sometimes pulse can be empty, therfore Eve can take advantage
of the multi-photon events and perform photon- number-splitting (PNS)
attack to keep some photons without being detected. To prevent this
PNS attack, several techniques have been made on weak laser pulses
such as a decoy-state method \cite{decoy} and SARG04 protocol \cite{SARG04_1},
which is robust against PNS attack. In decoy-state method, Alice inserts
decoy states (extra photon with different intensity) in the signal
states to create a confusion for Eve. Eve cannot distinguish that
a coming photon-state is a signal state or decoy state, confusion
creates an error and Alice and Bob will know the presence of Eve or
PNS attack. 

With the low cost or no extra work, the other robust method against
PNS attack is SARG04 protocol, which is proposed by Scarani, Acin,
Rivordy, and Gisin in 2004 \cite{SARG04_1}. SARG04 protocol is a
prepare-and-measure based QKD protocol and it is the modification
of the BB84 protocol. It is based on the four non-orthogonal states
similar as BB84, but the classical part is different. Several works
on SARG04 protocol have been proposed \cite{two_photon,SARG04,SARG041,SARG043,SARG042,BB84andSARG04,depolarizing,Decoy_statemethod}, few researchers compared the performance of BB84 with SARG04 protocol \cite{BB84andSARG04,depolarizing}.
The aim of this paper is, to study the effect of free-space losses
on SARG04 protocol and generates a highest secret key rate under free-space
losses. Here, we have considered geometrical and atmospheric losses
and used decoy-state method \cite{Decoy_statemethod} to achieve maximum
secret key rate for longer distances.

This paper is organized as follows: In Sec. \ref{sec:SARG04-Protocol},
SARG04 protocol is discussed in detail. The secret key generation
rate is discussed in Sec. \ref{sec:Secret-Key-Generation} and finally
in Sec. \ref{sec:Conclusion}, we have concluded the work.

\section{SARG04 Protocol \label{sec:SARG04-Protocol}}

Due to the lack of single-photon sources, weak coherent laser pulse
source is used in the single-photon based protocol which contains
sometimes more than one photon. Therefore, multi-photon pulses invites
PNS attack and Eve steals the information sent by Alice to Bob. For
that, SARG04 protocol is the best QKD protocol because it is robust
against PNS attack and it allows to generate a secret key not only
from single-photon pulses, but also two-photon pulses as well \cite{two_photon}.
In SARG04 protocol, Alice uses four sets of states $s_{1}=\left\{ |H\rangle,\,|45^{0}\rangle\right\} ,\,s_{2}=\left\{ |V\rangle,\,|45^{0}\rangle\right\} ,\,s_{3}=\left\{ |H\rangle,\,|-45^{0}\rangle\right\} ,\,s_{4}=\left\{ |V\rangle,\,|-45^{0}\rangle\right\} $
to publicly announce.

The SARG04 protocol goes as follows: 
\begin{enumerate}
\item Alice prepares a string of photons in one of four polarization states
$|H\rangle,$ $|V\rangle,$ $|45^{0}\rangle$ and $|-45^{0}\rangle$
and sends them to Bob by a quantum channel (optical-fiber, free-space
or underwater). 
\item Bob receives and randomly performs the measurement in $Z$$\left\{ |H\rangle,\,|V\rangle\right\} $
basis or $X$$\left\{ |45^{0}\rangle,\,|-45^{0}\rangle\right\} $
basis. 
\item In the classical part, here, Alice and Bob do not announce the basis,
Alice announces one of four sets $s_{1,\,}s_{2,\,}s_{3,\,}\text{or}\,s_{4}$
publicly that contain the state of the preparing photon. 
\item Bob shares with Alice to discard the times where the measurement output
is seems confusing. 
\item In the final step, they apply error correction method and privacy
amplification to generate the final secret key. 
\end{enumerate}
Quantum procedure of SARG04 is same as BB84, the only difference in
the classical procedure. In the SARG04 protocol, the secret key is
generated by single-photon as well as two-photon pulses, however in
BB84, the secret key is generated with only single-photon pulses.

.

\section{Secret Key Generation Rate\label{sec:Secret-Key-Generation}}

Alice and Bob generate a sifted key by performing some measurements
on the photon states, that sifted key may contain some errors due
to noise, losses and Eve. Therefore, they perform QBER test, some
error correction method and privacy amplification to know the errors
and reduce the knowledge of Eve in the sifted key. For SARG04 protocol,
secret key generation rate after privacy amplification with single-photon
and two-photon states is calculated by using the following formula,

\begin{equation}
\begin{array}{ccc}
SKR & = & -Q_{\mu}f\left(E_{\mu}\right)H_{2}\left(E_{\mu}\right)+Q_{1}\left[1-H_{2}(e_{1})\right]\\
 &  & +Q_{2}\left[1-H_{2}\left(e_{2}\right)\right]
\end{array}\label{eq:}
\end{equation}

Here are many terms to discuss in detail. Firstly, start with the
first term which is the fraction of EPR pairs spent for error correction.
The subscript $\mu$ represents the mean photon number. $Q_{\mu}$
and $E_{\mu}$ are the overall gain and overall quantum bit error
rate (QBER), respectively, for $n$-photon states. $f\left(E_{\mu}\right)$
is the error correction efficiency and $H_{2}(x)=-x\,\text{lo\ensuremath{g_{2}}}x-(1-x)\,\text{lo\ensuremath{g_{2}}}(1-x)$
is the Shannon entropy. Second and third terms are the contribution
to generate a secret key from the single-photon and two-photon states
for the SARG04 protocol. $Q_{1}$ and $Q_{2}$ are the gain and $e_{1}$
and $e_{2}$ are the bit error rate for the single-photon and two-photon
states. Now, we discuss the calculation to calculate the parameters
which cannot be directly estimated. Many papers have already been
discussed in detail \cite{SARG04,SARG041,SARG043,SARG042}, and proved
that the highest key rate with the longer distance can only be possible
with the help of decoy-state method \cite{Decoy_statemethod}. In
this paper, we have also used decoy-state method to calculate the
parameters such as yields, gain, error etc.

The transmission efficiency (overall transmittance) for an $n-$photon
signal is

\begin{equation}
\eta_{n}=1-\left(1-\eta\right)^{n},\label{eq:losses}
\end{equation}

where $\eta$ is the detection efficiency of Bob's side which includes
channel losses and optical losses such as, detector efficiency, coupling
efficiency, losses inside the detector box. In this study, free-space
losses: geometrical losses and atmospheric losses are considered as
channel loss

\begin{equation}
\eta=\eta_{Bob}.\left[\frac{d_{r}}{d_{t}+D.\text{\ensuremath{L}}}\right]^{2}.\text{exp}\left(-\alpha.\text{\ensuremath{L}}\right)\label{eq:-6}
\end{equation}

$d_{r}$ and $d_{t}$ are the diameters of receiver and transmitter
aperture, $D$ is the beam divergence and $\alpha$ is atmospheric
attenuation. 

The yield $Y_{n}$ is the conditional probability of a detection event
at Bob's side given that Alice emits an $n-$photon states, which
is \cite{SARG04}

\begin{equation}
Y_{n}=\eta_{n}\left(\frac{e_{det}}{2}+\frac{1}{4}\right)+\left(1-\eta_{n}\right)P_{dark}\frac{1}{2},\label{eq:-1}
\end{equation}

and the bit error rate for $n-$photon states is,

\begin{equation}
e_{n}=\frac{1}{Y_{n}}\left[\eta_{n}\frac{e_{det}}{2}+\left(1-\eta_{n}\right)P_{dark}\frac{1}{4}\right],\label{eq:-2}
\end{equation}

where, $P_{dark}$ is the background count and $e_{det}$ is the misalignment
in the detector and its characterizes the alignment and stability
of the optical system.

$Q_{n}$ is the gain of $n-$photon states which is calculated as,

\begin{equation}
Q_{n}=Y_{n}e^{-\mu}\frac{\mu^{n}}{n!},\label{eq:-3}
\end{equation}

In Eqs. \ref{eq:losses} \ref{eq:-1} \ref{eq:-2} \ref{eq:-3}, $n$
represents for $n-$photon signal states, however in this SARG04 protocol
key is generated with single-photon ($n=1$) and two-photon ($n=2$)
states. So, parameters $Y_{1},\,Y_{2,}\,e_{1},\,e_{2},\,Q_{1},$ and
$Q_{2}$ which are used in the generation of SKR is calculated by
putting $n=1\,\text{and}\,2$ in the above equations.

The overall gain $Q_{\mu}$(weighted average of the yields) and overall
QBER $E_{\mu}$(weighted average of the errors) for SARG04 protocol
is

\begin{equation}
Q_{\mu}=\frac{1}{2}P_{dark}e^{-\eta\mu}+\left(\frac{e_{det}}{2}+\frac{1}{4}\right)\left(1-e^{-\eta\mu}\right),\label{eq:-4}
\end{equation}

\begin{equation}
E_{\mu}=\frac{1}{Q_{\mu}}\left[\frac{1}{4}P_{dark}e^{-\eta\mu}+\frac{e_{det}}{2}\left(1-e^{-\eta\mu}\right)\right].\label{eq:-5}
\end{equation}

\begin{figure}
\begin{centering}
\includegraphics[scale=0.3]{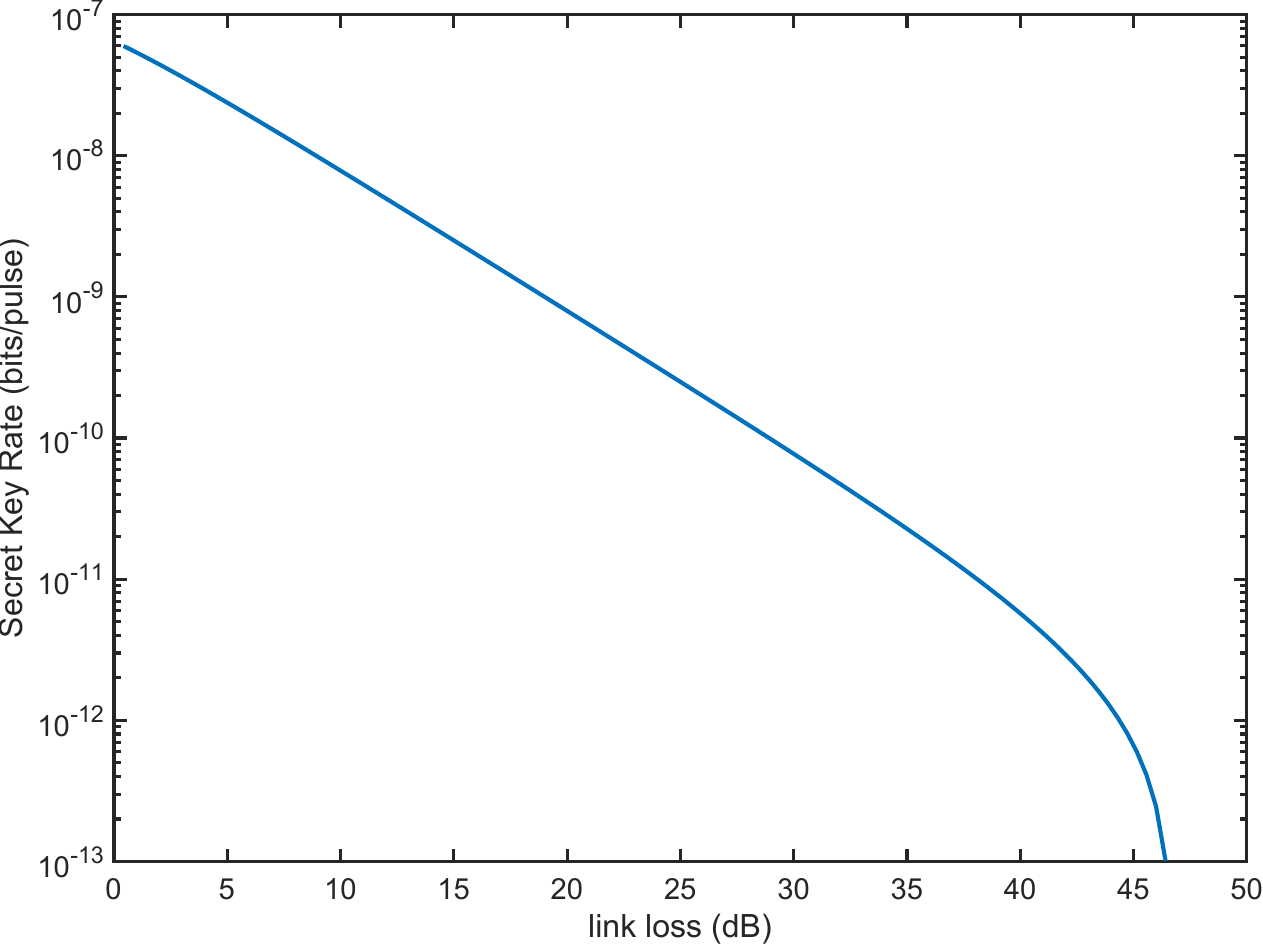} 
\par\end{centering}
\caption{\label{fig:Sceret-key-generation}Secret key generation rate in bits/pulse
under the effect of free-space losses: geometrical and atmospheric
losses. }
\end{figure}

In this study, the main task is the generation of highest SKR for
free-space based SARG04 QKD protocol with a maximum tolerance of noise.
For that, the main factor is the gain $Q_{n=1,\,2}$ of the signal
state which is directly dependent on the mean photon number $\mu.$
If $\mu$ is maximum than the gain is maximized, but the probability
of multi-photon states is also increased which invites eavesdropping.
So, $\mu$ is not to be too large and too small, it should optimal
mean photon number which helps to generate highest SKR and lowest
Eve attacks. In Ref. \cite{SARG04}, the variation of the different
values of $e_{det}$ to find the optimal value of $\mu$ with maximum
length is shown. To consider all the points in terms of the generation
of highest SKR, we have selected the values: $P_{dark}=10^{-6},$
$\eta_{Bob}=4.5\%,$ $e_{det}=0.033,$ $f(E_{\mu})=1.22$ $d_{r}=70\text{ mm},$
$d_{t}=10\text{ mm},$$D=0.025$ mrad \footnote{Above values are considered to minimize the losses in the generation
of SKR for the SARG04 protocol.} and the result is depicted in Fig. \ref{fig:Sceret-key-generation}.

\section{Conclusion\label{sec:Conclusion}}

SARG04 protocol generates a secret key with single-photon and two-photon
pulses. In this study, we have selected those optimal values which
is used to get the low bit error rate and increase the gain of the
signal states. Consequently, we have achieved highest secret key rate
with a maximum tolerance of noise$\sim47$dB under free-space losses.
Our results prove that with the selected values, we can implement
the SARG04 protocol for longer distances with the generation of the
highest secret key rate in free-space.

\textbf{Acknowledgment: }The authors thankfully acknowledge the following
funding agency: Defence Research and Development Organisation, India
(JATC-2 PROJECT $\#$ 2) for a project grant.

\end{document}